\documentclass[preprint]{aastex}
\usepackage{mathtext,bbm,amsmath,amsfonts,amssymb,indentfirst,syntonly,graphicx}
\usepackage[english]{babel}
\usepackage{slashbox}
\usepackage{calc}
\usepackage{tikz}
\usepackage{latexsym,graphicx,bbm}

\newcommand{\beq}{\begin{eqnarray}}
\newcommand{\eeq}{\end{eqnarray}}

\begin{document}

\title{Polarization of photons scattered by electrons in any spectral distribution}

\author{ Zhe Chang\altaffilmark{1,2}, Yunguo Jiang\altaffilmark{3,4,*}, Hai-Nan Lin\altaffilmark{1}}
\affil{\altaffilmark{1}Institute of High Energy Physics\\Chinese Academy of Sciences, 100049 Beijing, China}
\affil{\altaffilmark{2}Theoretical Physics Center for Science Facilities\\Chinese Academy of Sciences, 100049 Beijing, China}
\affil{\altaffilmark{3} School of Space Science and Physics \\Shandong University at Weihai,  264209 Weihai, China}
\affil{\altaffilmark{4} Shandong Provincial Key Laboratory of Optical Astronomy \\ and Solar-Terrestrial Environment, 264209 Weihai, China}
\altaffiltext{*}{On leave of absence from IHEP, Beijing. jiangyg@ihep.ac.cn}

\begin{abstract}

  Based on the quantum electrodynamics, we present a generic formalism of  the polarization for beamed monochromatic photons scattered by electrons in any spectral distribution. The formulae reduce to the components of the Fano matrix when electrons are at rest. We mainly investigate the polarization in three scenarios, i.e., electrons at rest, isotropic electrons with a power law spectrum and thermal electrons. If the incident beam is polarized, the polarization is reduced  significantly by isotropic electrons at large viewing angles, and the degree of polarization due to thermal electrons is about one times less than that of electrons in a power law. If the incident bean is unpolarized, soft $\gamma$-rays can lead to about $15\%$ polarization at viewing angles around $\pi/4$. For isotropic electrons,  one remarkable feature is that the polarization as a function of the incident photon energy always peaks roughly at $1$ MeV, this is valid for both the thermal and power law cases. This feature can be used to distinguish the model of the inverse Compton scattering  from that of the synchrotron radiation.

\end{abstract}

\keywords{polarization - radiation mechanism: non-thermal - scattering}

\section{Introduction}

Synchrotron radiation (SR) and inverse Compton scattering (ICS) are considered most frequently as the main mechanisms of producing the X-ray and $\gamma$-ray in astrophysical objects such as the gamma-ray burst (GRB), the active galactic nuclei (AGN) and blazars.  Besides light curves and spectra, the polarization is another important tool for the diagnosis of the emission mechanisms of the astrophysical objects, since the degree of polarization, orientation, and time variation can provide more information for analysis. Many  models have been suggested to investigate the polarization of SR and ICS in different setups. The polarization of SR with power law distributed electrons in a uniform magnetic field is well-known, i.e.,  $\Pi=(p+1)/(p+7/3)$, where $p$ is the spectral index of electrons \citep{Westfold:1957,Ginzburg:1965,Ginzburg:1969}.  The investigation of ICS induced polarization is still in progress.

\citet{Bonometto:1970}  presented a general formalism of the polarization in ICS for arbitrary distributions of photons and electrons in the Thomson limit. Later on, \cite{Bonometto:1973} used the same scenario to investigate the polarization for  ICS with the SR source, they found that the initial polarization is preserved with a large fraction after ICS.
\citet{Begelman:1987} investigated the polarization in a system that the ambient isotropic radiation is scattered by the flow of cold electrons.  Different geometrical structures of the electron jet (e.g., pencil beam or hollow cone) produce  polarizations with different orientations. The initial ambient radiation is unpolarized, but the maximal $100\%$ polarization can be achieved at the  viewing angle $\theta \sim 1/\gamma_e$, where $\gamma_e$ is the Lorentz factor of electrons. \citet{Celotti:1994} used the similar method to study the polarization properties of synchrotron self-Compton (SSC) emission for an isotropic electron distribution, their results roughly agree with that of \citet{Bonometto:1973}.

\citet{Aharonian:1981} considered the spectrum and polarization due to ICS. The formula of the polarization was given for the unpolarized initial beam. Numerical simulation for the isotropic electron distribution shows that a moderate linear polarization is still possible at large viewing angles \citep{Fan:2009}. \citet{McNamara:2009} investigated the polarization properties of X-rays in the AGN. Their simulations show that the polarization increases with the viewing angle for source photons emitted from an accretion disc or from the cosmic microwave background (CMB), and the maximal polarization is about $20\%$.  For the SSC case, they  found that the resulting X-ray polarization depends
strongly on the injection site of the seed photon, i.e.,  with typical $\Pi \approx 10 \sim 20 \%$ for the emission  near the jet base, while $\Pi \approx 20 \sim 70 \%$  for the uniform emission throughout the jet.

GeV photons are observed in the prompt phase in some GRBs \citep{Abdo:2009,Ackermann2010}. Although the photon energy is much reduced in the jet frame by excluding the Lorentz effect, MeV photons are still popular in the jet frame in GRBs. If these photons are the results of ICS,  the formalism of the polarization  should be extended to  the KN regime. Another shortcoming of these investigations is that the incident photons are unpolarized except SSC.
\citet{Krawczynski:2012} used the Monte Carlo simulation to calculate the polarization of ICS in the KN regime for isotropic electrons, and found that the net polarization is zero, that is against the results given by \citet{McNamara:2009}. We  presented a formalism of the polarization induced by cold electrons \citep{Chang:2013}. We found that the orientation of the polarization  changes $90^{\circ}$ when the incident photons are of the SR origin, and the polarization reduces significantly in the KN regime. However, an analytical formalism  of the polarization in the ICS process with arbitrary initial polarizations and electron spectra is still lacking.

In the present paper, we present an analytical formalism for the polarization induced by isotropic electrons with an arbitrary spectrum.  First we consider the kinematics of the scattering of a photon by an electron. The cross section for the scattering of polarized photons by an unpolarized electron is used to obtain the  Stokes parameters for the outgoing photon. In the framework of the quantum electrodynamics, we investigate the polarization effects  induced by the isotropic electrons for both the polarized and unpolarized incident photons.When electrons are at rest, the components of the cross section reduce to the components of the Fano matrix. By numerical calculations, we discuss the polarization for electrons with power law and thermal distributions, respectively. The rest of the paper is arranged as follows. In Section \ref{sec:setup}, we introduce the kinematics of the scattering, and present the formulae of the Stokes parameters for the scattered photon. In Section \ref{sec:result}, we investigate the polarization for  three scenarios. Discussion and conclusions are given in Section \ref{sec:DC}.

\section{Polarization formula } \label{sec:setup}

\begin{figure}
\centering
  \plotone{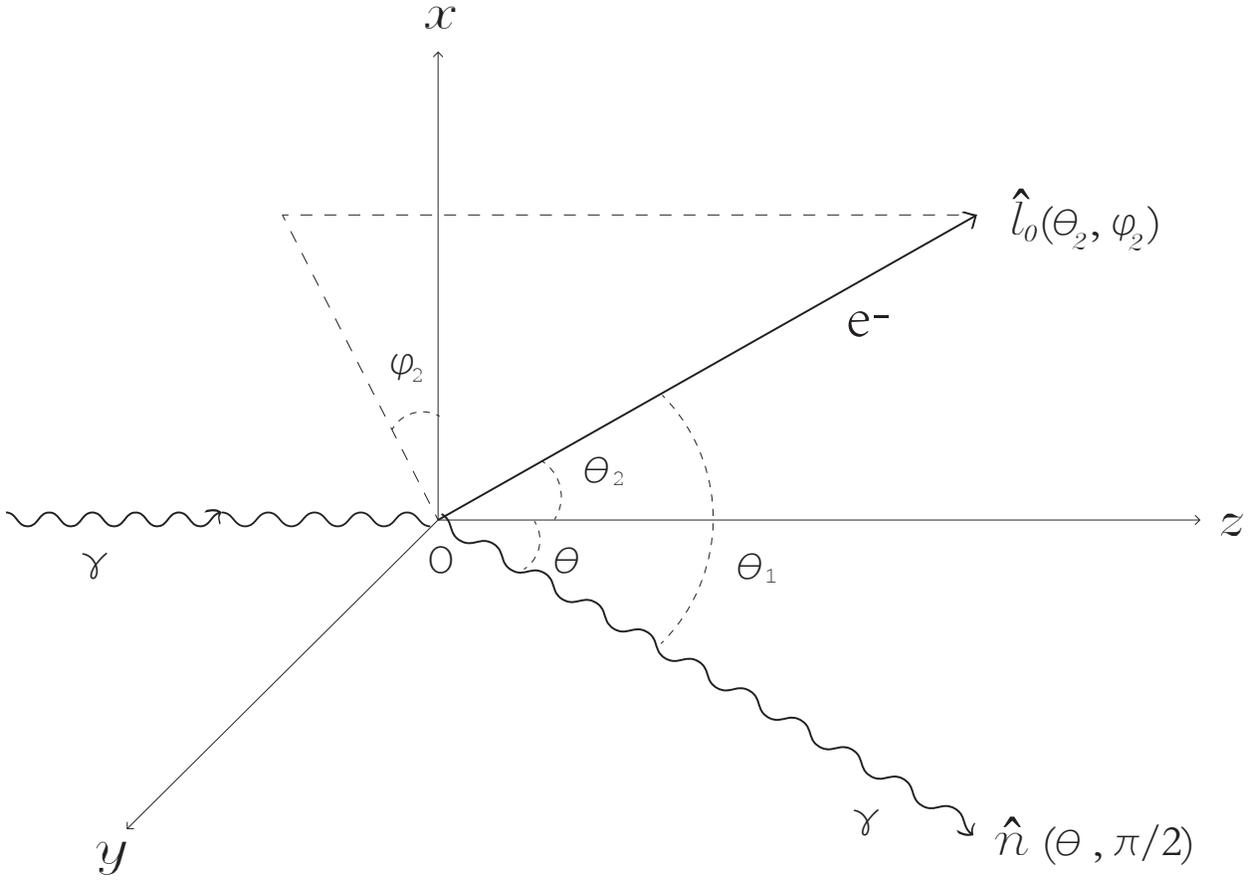}
  \caption{Schematic representation of the scattering process in the jet comoving frame. The incident photon goes along the positive $z$ direction, scattered with an electron  at point $O$, then moves along the line of sight $\hat{n}$. The initial electron is injected along the $\hat{l}_0$ direction, and the moving direction of the scattered electron is ignored here. The polar and azimuthal angles of $\hat{l}_0$ and $\hat{n}$ are ($\theta_2, \varphi_2$) and ($\theta, \pi/2$), respectively. The angle between $\hat{l}_0$ and $\hat{n}$ is denoted by $\theta_1$. } \label{fig:geo}
\end{figure}

The scattering process is illustrated in Fig. \ref{fig:geo}. The beam of photons produced by the synchrotron radiation of electrons is collimated roughly along the radial direction of the outflow, namely, the $z$-direction. Suppose that a photon is injected along $\hat{z}$ (we take the convention that a variable with a hat denotes the unit vector along that direction), it is scattered by an electron at point $O$.  Then, it goes toward the observer along the $\hat{n}$ direction.  The $y$-axis is chosen in the $\hat{z}$ and $\hat{n}$ plane, and the axes $xyz$ form a right-handed set.  The injected electron has an arbitrary momentum $\vec{p}_0= \gamma \beta m_e c \hat{l}_0$, where $\gamma$ is the Lorentz factor of the electron, $\beta=|v|/c$, and $\hat{l}_0$ denotes the  incident direction of the electron. The momentum of the injected electron can be described by  spherical coordinates $(\theta_2, \varphi_2)$, and $\gamma$ completely.  $\theta_2$ is the polar angle between $\hat{z}$ and $\hat{l}_0$, and $\varphi_2$ is the azimuthal angle of $\hat{l}_0$ relative to the $x$-axis, measured counter clockwise from the observer side.

 The four momenta of the electron before and after scattering process are  written as $P^0_{e}=(E_0/c,\vec{p}_0)$ and $P^1_{e}=(E_1/c,\vec{p}_1)$, respectively. In a similar way, the four momenta of the photon before and after scattering are written as $P^0=\epsilon_0/c(1,\hat{z})$ and $P^{1}=\epsilon_1/c(1,\hat{n})$. The conservation of the energy and momentum of the system states that
\begin{equation} \label{eq:emconservation}
P^0_{e}+P^0=P^1_{e}+P^1.
\end{equation}
After scattering, the photon energy is written as
\begin{equation} \label{eq:epspr}
\epsilon_1=\frac{\epsilon_0(1-\beta \hat{z}\cdot \hat{l}_0)}{\frac{\epsilon_0}{\gamma m_e c^2}(1-\hat{z}\cdot \hat{n})+1-\beta \hat{l}_0 \cdot \hat{n}}.
\end{equation}
The angle between $\hat{z}$ and $\hat{n}$ is denoted by $\theta$, and the angle between $\hat{l}_0$ and $\hat{n}$ is denoted by $\theta_1$. The geometrical relation gives
\beq \label{eq:angrel}
\cos \theta_1 = \cos \theta \cos \theta_2 + \sin \theta \sin \theta_2 \sin \varphi_2.
\eeq
 One can also calculate the Lorentz factor of the electron after scattering. However, they are not observable quantities and have no effects on the final polarization.

In quantum electrodynamics, a convenient way to describe the partial polarization is by using the density matrix $\rho$ in terms of three real Stokes parameters $\xi_i$ ($i=1,2,3$). The density matrix is written as \citep{Berest:1982}
\beq  \label{eq:rho}
\rho = \frac{1}{2}\left( \begin{array}{cc}
 1+\xi_3 & \xi_1+i \xi_2 \\
\xi_1-i \xi_2 & 1-\xi_3  \end{array} \right),
\eeq
where the Stokes parameters $\xi_i$ are defined with respect to the axes $xyz$.
 The positive (negative)  $\xi_3$ describes that the photon is linearly polarized along the $x$ ($y$) axis,
the parameter $\xi_1$ means the linear polarization along the directions with azimuthal angles $\pm\pi/4$ relative to the $x$ axis in the $xy$ plane, and the parameter $\xi_2$ describes the left-hand or the right-hand circular polarization.
In terms of  Stokes parameters $\xi_i$, the polarization degree  is written as
\beq \label{eq:pi}
\Pi = \sqrt{\xi_1^2+\xi_2^2+\xi_3^2}.
\eeq
The Stokes parameters satisfy the condition $0\leq \xi_1^2+\xi_2^2+\xi_3^2 \leq 1$. For $\xi_i=0$, the photon is in a unpolarized state; for $\sum_{i=1}^{3} \xi_i^2=1$, the photon is completely polarized. In the $0< \Pi < 1$ case, we call that the photon is partially polarized.

When the $x$ and $y$ axes are rotated by an angle $\varphi$ about the $z$ axis, one obtains a new $x'y'z$ coordinate frame. Respect to these new  axes, the Stokes parameters  $\xi_i'$ transform as
\begin{align} \label{eq:xitra}
\xi'_1&=\xi_1 \cos 2 \varphi - \xi_3 \sin 2 \varphi,  \nonumber \\
 \xi'_2&= \xi_2,  \nonumber \\
\xi'_3&= \xi_3 \cos 2 \varphi + \xi_1 \sin 2 \varphi.
\end{align}
Therefore, the circular polarization is unaffected by the rotation, while the values of $\xi_1$ and $\xi_3$ depend on the chosen of the axes. However, one can easily verify that the degree of the polarization $\Pi$ is invariant under such rotation \citep{Mandel:1995}.
 Another alternative manner to describe the polarization is by using four Stokes parameters $I$, $Q$, $U$, and $V$ \citep{Rybicki:1979}, where $I$ is the total energy density of the photon. The three Stokes parameters $\xi_i$ correspond to $Q$, $U$ and $V$ normalized by $I$, i.e., $\xi_3=Q/I$, $\xi_1=U/I$ and $\xi_2=V/I$, respectively.

The differential cross section for the scattering of a polarized photon by an unpolarized electron is written as \citep{Berest:1982}
\begin{align} \label{eq:cs}
d \sigma = & \frac{1}{4} r_e^2 \left(\frac{\epsilon_1}{\epsilon_0}\right)^2 \left( \frac{\epsilon_1}{\epsilon_0}+\frac{\epsilon_0}{\epsilon_1}-\sin^2 \theta \right) \sin \theta d \theta d \varphi\nonumber \\
& + 2 r_e^2 \frac{d y }{x^2}d \varphi \bigg[ \xi_1 \xi_1'
\left( A+\frac{1}{2} \right)+\frac{1}{4}\xi_2\xi'_2   \left( 1 + 2A \right) B \nonumber \\
&+\xi_3\xi'_3 \left( A^2+ A+\frac{1}{2}\right)-(\xi_3+\xi'_3) \left( A^2+ A \right) \bigg],
\end{align}
where
\beq \label{eq:ab}
A \equiv \frac{1}{x}-\frac{1}{y}, \qquad B \equiv \frac{x}{y}+\frac{y}{x},
\eeq
and $r_e$ is the classical electron radius. The  Stokes parameters $\xi_i$ and $\xi'_i$ ($i=1, 2, 3$) describe the  polarizations of the incident and scattered photons, respectively. The parameters $x$ and $y$ are defined as
\begin{align} \label{eq:xy}
x\equiv \frac{2 P^0_e P^0}{m_e^2 c^2}&=\frac{2\gamma \epsilon_0}{m_e c^2}(1-\beta \cos \theta_2), \nonumber \\
y \equiv \frac{2P^0_e P^1}{m_e^2 c^2}&=\frac{2\gamma \epsilon_1}{m_e c^2}(1-\beta \cos \theta_1).
\end{align}
The dimensionless  variables $x$ and $y$ originate from the Mandelstam variables, and are always positive according to this expression.

In the setup, we suppose that the injected electrons are isotropic in the jet frame. If the initial momenta $P_e^0$ and $P^0$ are given, the momentum of the scattered photon is a function of $\theta$. Therefore, the differential of $y$ is written as
\beq \label{eq:dy}
dy=\frac{2 \epsilon_1^2}{m_e^2c^4}\left( 1-\frac{\beta \sin \theta_2 \sin \varphi_2}{1-\beta \cos \theta_2} \tan \frac{\theta}{2}\right) \sin \theta d \theta.
\eeq
Taking use of Equations (\ref{eq:epspr}), (\ref{eq:angrel}), (\ref{eq:cs}), (\ref{eq:xy}) and (\ref{eq:dy}), we rewrite the differential cross section  as
\begin{align}\label{eq:cs1}
d \sigma = & \frac{1}{4} r_e^2 d \Omega \left(\frac{\epsilon_1}{\epsilon_0}\right)^2  \bigg[ F_0 + F_{11} \xi_1 \xi_1' +F_{22} \xi_2\xi'_2\nonumber \\
& +F_{33} \xi_3\xi'_3 +F_3(\xi_3+\xi'_3)\bigg],
\end{align}
where $d\Omega=\sin \theta d \theta d \varphi$, and
\begin{align} \label{eq:fs}
F_0 \equiv & \frac{\epsilon_1}{\epsilon_0}+\frac{\epsilon_0}{\epsilon_1}-\sin^2 \theta ,\nonumber \\
F_{11}\equiv  &(A+\frac{1}{2}) \Sigma,  \nonumber \\
F_{22}\equiv & \frac{1}{4}(1+2A)B \Sigma, \nonumber \\
F_{33}\equiv & (A^2+A+\frac{1}{2}) \Sigma, \nonumber \\
F_3 \equiv &-(A^2+A) \Sigma,  \nonumber \\
\Sigma \equiv & \frac{4\left(1-\frac{\beta \sin \theta_2 \sin \varphi_2}{1-\beta \cos \theta_2} \tan \frac{\theta}{2} \right)}{\gamma^2 (1-\beta \cos \theta_2)^2}.
\end{align}
The Stokes parameters of the secondary photon are denoted by $\xi^{\rm f}_i$, which are equal to the ratios of the coefficients of  $\xi'_i$ to the terms independent of $\xi'_i$ in Equation (\ref{eq:cs1}) \citep{Berest:1982}, i.e.,
\begin{align} \label{eq:pol}
\xi^{\rm f}_1=& \frac{ \xi_1 F_{11}}{F_0+\xi_3 F_3}, \nonumber \\
 \xi^{\rm f}_2=& \frac{ \xi_2 F_{22}}{F_0+\xi_3 F_3}, \nonumber \\
\xi^{\rm f}_3=& \frac{F_3+ \xi_3 F_{33}}{F_0+\xi_3 F_3}.
\end{align}
Noticing that the circular polarization of the scattered photon occurs only if the incident photon is circularly polarized.
The degree of polarization of the scattered photon can be expressed as
\beq \label{eq:polfinal}
\Pi=\sqrt{(\xi^{\rm f}_1)^2+(\xi^{\rm f}_2)^2+(\xi^{\rm f}_3)^2}.
\eeq

We consider that the incident photon beam is of the SR origin. Suppose that there is a uniform and static magnetic field $\vec{B}$ in the plasma, the azimuthal angle of $\vec{B}$ projected on the $xy$ plane is denoted by $\varphi_0$.  The definition of the polarization  is $\Pi_0\equiv(I_2-I_1)/(I_2+I_1) $, where  $I_2$ and $I_1$ are the intensities along the direction $\hat{z} \times \vec{B}$ and $(\hat{z} \times \vec{B}) \times \hat{z}$ with the azimuthal angle $\varphi_0+\pi/2$ and $\varphi_0$, respectively. The primary degree of the polarization  of SR photons is $\Pi_0= (p+1)/(p+7/3)$. The Stokes parameters of the incident photons can be written as
\begin{align} \label{eq:xi0}
\xi_1=&\Pi_0 \sin 2 \varphi_0, \qquad
\xi_2=0, \qquad
 \xi_3= \Pi_0 \cos 2 \varphi_0.
\end{align}
When $\varphi_0=0$ (the magnetic field is perpendicular to the scattering plane), one has $\xi_1=\xi_2=0$ and $\xi_3=\Pi_0$. The orientation of the polarization is along the $x$ axis. When $\varphi_0=\pi/4$, one has $\xi_1=\Pi_0$ and $\xi_3=0$. The orientation of the polarization is along the line rotating  $\pi/4$  counter clockwise from the $x$ axis. Therefore, $\varphi_0$ represents the initial polarization orientation.
 After being scattered by electrons, the direction of the polarization changes. $\xi_i^{\rm f}$ are defined in a new coordinate axes, i.e., the  $x$ axis is fixed and  $z$ axis is rotated to coincide with the direction $\hat{n}$. Then, $\xi^{\rm f}_3$ still describes the polarization along the $x$ axis, and $\xi^{\rm f}_1$ describes the polarization along the line with azimuthal angle $\pi/4$ in the new set of axes.

\section{Effects of different electron spectra} \label{sec:result}

The obtained final polarization is quite general. It depends on the spectra of  incident photons and injected electrons.  We investigate the polarization in three scenarios: electrons at rest, isotropic electrons with a power law distribution and electrons with the Maxwell distribution. For electrons at rest, we will recover the results obtained in \citet{Chang:2013}. Electrons with the thermal and the power law distributions are common populations in the astrophysical sources.

\subsection{Electrons at rest}

First we investigate a simplified system, in which the injected electrons are at rest in the jet frame. Such system has already been studied in detail \citep{Chang:2013}. We will show that it is a special case of the setup in the present paper. Since electrons are static,  i.e., $\beta=0$ and $\gamma=1$, the scattered photon energy is written as
\beq \epsilon_1= \frac{\epsilon_0 }{1+ \frac{\epsilon_0}{m_e c^2}(1- {\rm cos}\theta)}.
\eeq
 Formulae in Equation (\ref{eq:fs}) reduce to \citep{Berest:1982}
\begin{align} \label{eq:fsta}
F_0=& \frac{\epsilon_1}{\epsilon_0}+\frac{\epsilon_0}{\epsilon_1}-\sin^2 \theta , \nonumber \\
F_3=&  \sin^2 \theta, \nonumber \\
F_{11}=&2 \cos \theta, \nonumber \\
 F_{22}=& \left(\frac{\epsilon_1}{\epsilon_0} +\frac{\epsilon_0}{\epsilon_1} \right) \cos \theta , \nonumber \\
 F_{33}=&1+ \cos^2 \theta.
\end{align}
These expressions are also components of the Fano matrix \citep{Fano:1957}.
For an initially linearly polarized photon, the Stokes parameters are  given in Equation (\ref{eq:xi0}). The final Stokes parameters are written as
\begin{align} \label{eq:popost}
 \xi^{\rm f}_1=&\frac{2 \Pi_0 \cos \theta  \sin 2\varphi_0}{\epsilon_1/\epsilon_0+\epsilon_0/\epsilon_1-(1-\Pi_0 \cos 2 \varphi_0)\sin^2 \theta}, \nonumber \\
  \xi_2^{\rm f}=&0, \nonumber \\
\xi^{\rm f}_3=&\frac{\sin^2 \theta+ \Pi_0 \cos 2\varphi_0(1+\cos^2 \theta)}{\epsilon_1/\epsilon_0+\epsilon_0/\epsilon_1-(1-\Pi_0 \cos 2\varphi_0)\sin^2 \theta}.
\end{align}
When the polarization direction is parallel to the scattering plane, i.e., $\varphi_0=\pi/2$, the expression of $\xi^{\rm f}_3$ agrees with the formula given by \citet{Chang:2013}. A remarkable feature of the polarization in such a setup is that the direction of polarization changes $90^{\circ}$ after the scattering process. \citet{McNamara:2009} used a similar formula to calculate the X-ray polarization in AGN, except that the incident photons are completely polarized, i.e., $\Pi_0=1$.

\begin{figure}
\centering
  \plotone{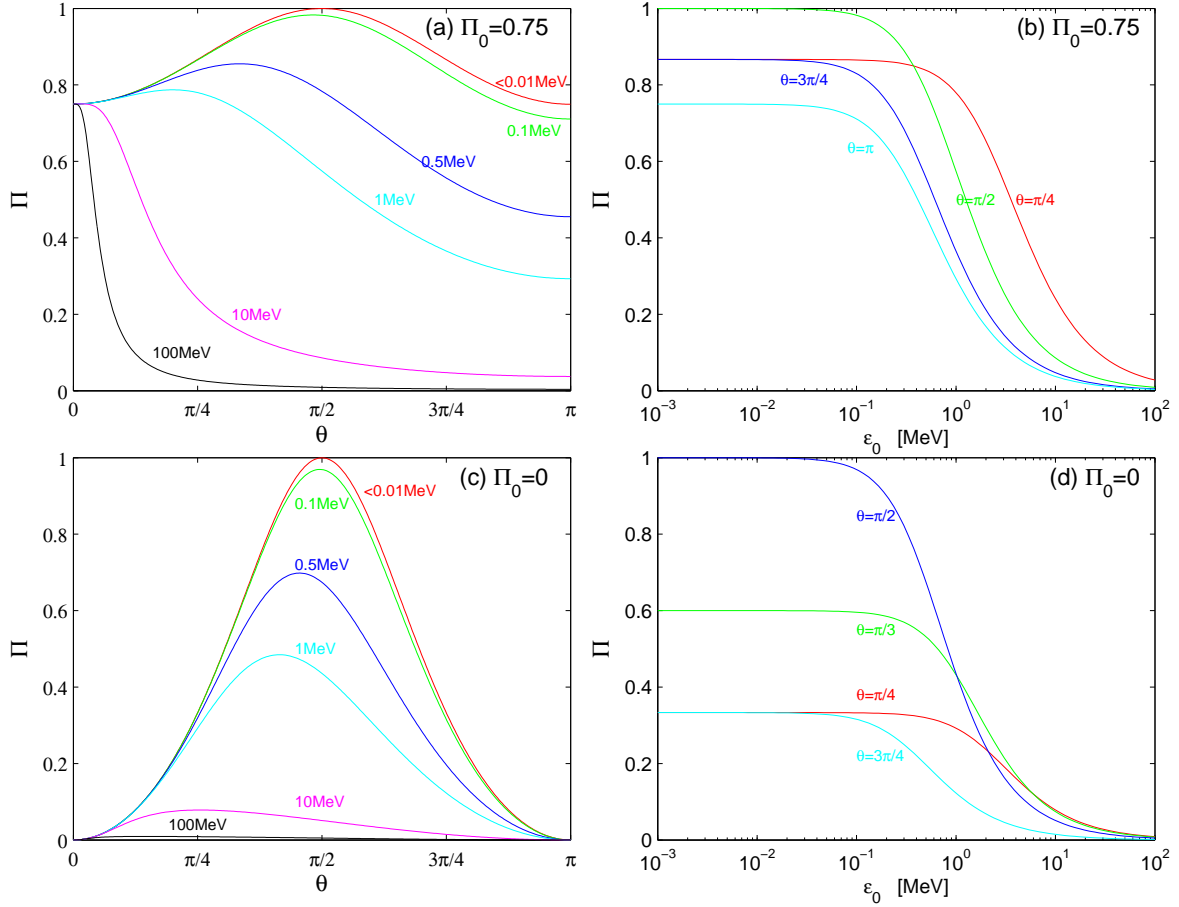}
  \caption{Polarization as a function of the incident photon energy $\epsilon_0$ and the scattering angle $\theta$ for the static electrons. The upper two panels depict that the incident photons are polarized with degree of polarization $\Pi_0=0.75$. The lower two panels depict the unpolarized incident radiation. In the left two panels, the polarization as a function of $\theta$ is plotted for different energies $\epsilon_0=0.01$, $0.1$, $0.5$, $1$, $10$ and $100$ MeV, respectively. The curve for the $0.001$ MeV photon almost coincides with the curve of the $0.01$ MeV photon, so we use the  ``$<0.01$ MeV" symbol to denote that. In the right two panels, polarization as a function of $\epsilon_0$ is plotted for different viewing angles $\theta=\pi/4$, $\pi/3$, $\pi/2$, $3\pi/4$ and $\pi$, respectively.} \label{fig:pirest}
\end{figure}

In Fig. \ref{fig:pirest}, we plot the polarization as a function of $\theta$ and $\epsilon_0$ for the static electron case. Since injected electrons are at rest, the Lorentz factor  $\gamma$ is equal to one, and the Stokes parameters are independent of $\theta_2$, $\varphi_2$. For all the four panels in Fig.\ref{fig:pirest}, we set $\varphi_0=0$. The top two panels show the case that the initial polarization is $\Pi_0=0.75$. From the left top panel, it is evident that the polarization achieves $100\%$ at $\theta=\pi/2$ for small energy photons $\epsilon_0=0.01$ MeV. This agrees with the classical Thomson limit. When $\epsilon_0$ increases, the maximal polarization decreases. For instance, the maximum degree of the polarization is almost $\Pi_0$ for $\epsilon=1$ MeV.  For large energy photons, $\Pi$ goes to zero  quickly when $\theta$ becomes large. This is due to the KN effect. A more evident manifestation of the KN effect is shown in the right top panel. At different viewing angles  $\theta=\pi/4$, $\pi/2$, $3\pi/4$ and $\pi$, the degree of the  polarization starts to decrease at $\epsilon_0 \approx 1$ MeV, and tends to zero at $100$ MeV.
The bottom two panels depict the polarization for the unpolarized incident photons. The left bottom one shows that low energy photons have larger degree of polarization than the high energy photons. Photons with energy higher than $10$ MeV have small polarization. The maximum polarization occurs at $\theta \approx \pi/2$, this is the same with the initially polarized case. The right bottom panel shows how $\Pi$ tends to zero as the energy increasing.

\subsection{The power law distribution}

 We aim to obtain the polarization when the monochromatic incident radiation is scattered by isotropic electrons. We consider the case that the incident beam is linearly polarized ($\xi_2=0$), and the orientation of the polarization is arbitrary. We denote the spectrum of the electrons by   ${\cal N(\gamma)}$. The scattered beam has an axial symmetry relative to $z$ axis, so that $d\sigma$ is independent of the azimuthal angle $\varphi$. After averaging over the spectrum and angles of  initial electrons, we obtain the averaged cross section  as
\begin{align}\label{eq:csa}
\big\langle \frac{d \sigma }{d\Omega}(\epsilon_0,\theta) \big\rangle =&\frac{\int_{\gamma_1}^{\gamma_2}{\cal N(\gamma)} d\gamma \int_0^{4\pi}  d \Omega(\theta_2,\varphi_2) \frac{d \sigma}{d \Omega}  }{ 4 \pi \int_{\gamma_1}^{\gamma_2}{\cal N(\gamma)} d\gamma } \nonumber \\ =&  \frac{1}{4} r_e^2  \bigg[ \langle F_0 \rangle  +\langle F_{11} \rangle \xi_1 \xi_1'+\langle F_{22} \rangle \xi_2\xi'_2 \nonumber \\
&\phantom{ \frac{1}{4} r_e^2  \bigg[ } +\langle F_{33} \rangle \xi_3\xi'_3+ \langle F_3 \rangle (\xi_3+\xi'_3) \bigg].
\end{align}
The definitions of averaged components  $\langle F_a \rangle$ ($a=$  $0$,  $11$, $22$, $33$, 3) are similar to the averaged cross section, i.e.,
\beq \label{eq:avf}
\langle F_a (\epsilon_0,\theta)\rangle=  \frac{\int_{\gamma_1}^{\gamma_2}{\cal N(\gamma)} d\gamma\int_0^{4\pi}  d \Omega(\theta_2,\varphi_2)  \left( \frac{\epsilon_1}{\epsilon_0}\right)^2 F_a}{4\pi \int_{\gamma_1}^{\gamma_2}{\cal N(\gamma)} d\gamma}.
\eeq
Here the $\epsilon_1^2/\epsilon_0^2$ term is included, because the Stokes parameters are the ratio of intensities which are proportional to the cross section.
The integral on the electron momentum is constrained by a physical condition \citep{Kibble:1960}, which is written as $y \geq x/(1+x)$.
However, this condition constrains only a small region in the phase space. This constrain can be ignored when  the integration is performed.
 The final Stokes parameters can be obtained by replacing the components $F_a$  with the averaged components $\langle F_a \rangle$, i.e.,
\begin{align}\label{eq:pola}
  \xi^{\rm f}_1= &\frac{ \xi_1 \langle F_{11} \rangle }{\langle F_0 \rangle  +\xi_3 \langle F_3 \rangle  }, \nonumber \\
 \xi^{\rm f}_2= &\frac{ \xi_2\langle F_{22} \rangle }{\langle F_0 \rangle +\xi_3 \langle F_3 \rangle }, \nonumber \\
 \xi^{\rm f}_3= &\frac{\langle F_3 \rangle + \xi_3 \langle F_{33}\rangle  }{\langle F_0 \rangle  +\xi_3 \langle F_3\rangle  }.
\end{align}
It is difficult to give an analytical expression of these components. Numerical method is applied to estimate the final polarization. After integrating out the primary electron momentum variables $\gamma$, $\theta_2$ and $\varphi_2$, we obtain the degree of the polarization $\Pi$ as a function of the scattering angle $\theta$ and the incident photon energy $\epsilon_0$.

\begin{figure}
\centering
  \plotone{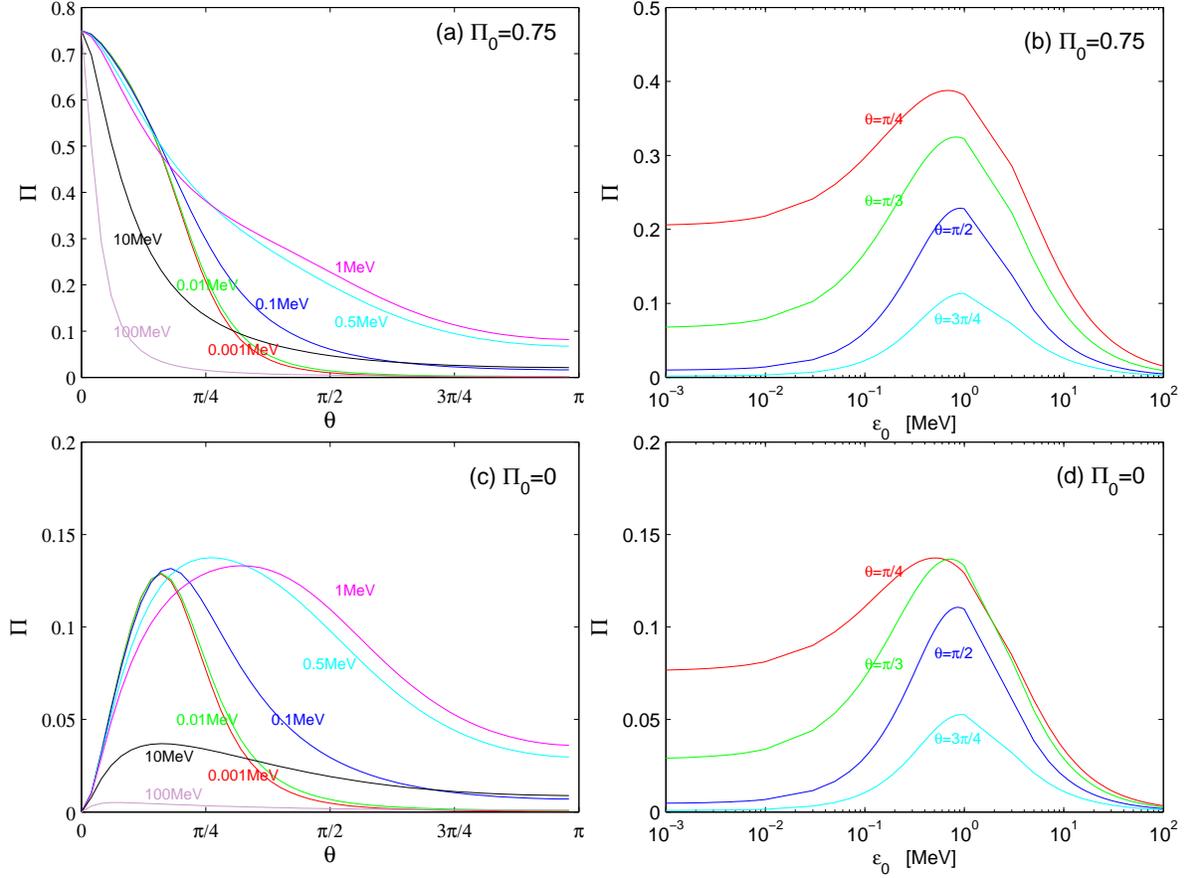}
  \caption{Polarization as a function of  $\epsilon_0$ and  $\theta$ for the isotropic electrons with a power law distribution, ${\cal N}(\gamma) \propto \gamma^{-3}$. In the top two panels, the incident photons are of SR origin and $\Pi_0=0.75$. The bottom two panels show the polarization with unpolarized incident radiation. The left two panels plot the polarization as a function of $\theta$ for different energies $\epsilon_0=0.001$, $0.01$, $0.1$, $0.5$, $1$, $10$ and $100$ MeV, respectively. The right two panels plot the polarization as a function of $\epsilon_0$  for different viewing angles $\theta=\pi/4$, $\pi/3$, $\pi/2$ and $3\pi/4$, respectively.  } \label{fig:pipower}
\end{figure}

By  numerical method, we estimate the polarization for the isotropic distributed electrons. The spectrum of injected electrons is a power law, i.e., ${\cal N}(\gamma)\propto \gamma^{-3}$ \footnote{Other spectral indices are allowed. Here we take $p=3$ for a example}. Due to the power law distribution, the range of integration on higher values of $\gamma$ are strongly suppressed. In the numerical calculation, we take the most significant contribution in the range $1 \leq \gamma \leq 10$.   The final polarization is tested for different values of $\varphi_0$, no significant variation is observed. Thus, we may set the initial azimuthal angle to be zero. Since electrons are isotropic, the integration ranges of $\theta_2$ and $\varphi_2$ are taken to be $0 \sim \pi$ and $0 \sim 2 \pi$, respectively. With these setups, the degree of polarization as a function of $\epsilon_0$ and $\theta$ are plotted in Fig. \ref{fig:pipower}.

In the top two panels, the incident radiation is polarized, and the polarization degree is $\Pi_0=0.75$.
We plot $\Pi$ as a function of $\theta$ for various energies in the top left panel. One finds that the maximal $\Pi$ is just $\Pi_0$ at $\theta=0$, where the incident photons are not scattered. A common feature of curves with different energies is that they all decrease at large viewing angle $\theta$, which means the isotropic distributed electrons reduce the polarization significantly. For $\epsilon_0=100$ MeV, $\Pi$ goes to zero even for $\theta < \pi/4$. Comparing with the left top panel in Fig. \ref{fig:pirest}, we conclude that the $100\%$ polarization cannot be realized in the isotropic case.  The curves of $0.001$ and $0.01$ MeV almost coincide, because they both are close to the Thomson limit. One also observes that curves of $0.001$, $0.01$ and $0.1$ MeV cross with curves of $0.5$, $1$ and $10$ MeV. For viewing angle $\theta> \pi/2$, $1$ MeV photons have the largest degree of polarization. This special feature is more evident in the top right panel. We plot $\Pi$ as a function of $\epsilon_0$ for viewing angles $\theta=\pi/4$, $\pi/3$, $\pi/2$ and $3\pi/4$, respectively. All these four curves have a maximal value around $\epsilon_0 \approx 1$ MeV, and the maximal value decreases as $\theta$ increasing. For instance, the maximum  polarization is about $40\%$ for the $1$ MeV photon at $\theta\approx\pi/4$. The initial SR induced polarization is mainly reserved at small viewing angles $\theta < \pi/4$, and is less than $40\%$ for viewing angles $\theta>\pi/4$. This result agrees with the polarization in the SSC processes \citep{Celotti:1994}.

The bottom two panels depict the initially unpolarized case. The left bottom panel plots $\Pi$ as a function of $\theta$ for different energies. The $100$ MeV photons have almost no polarization, and $10$ MeV photons  have polarization smaller than $4\%$. The maximum polarizations of all curves  occur at $\theta<\pi/2$, and are no more than $15\%$. Therefore, the high linear polarization is difficult to realize for unpolarized photons scattered by isotropic electrons. For viewing angles $\theta> \pi/2$, the degree of polarization of $1$ MeV photons is larger than that of photons with other energies. The highest value of $\Pi$ is achieved at $\theta \approx \pi/4$ for $\epsilon_0 \approx 0.5$ MeV. In the right bottom panel, we show how $\Pi$ varies with the energy for   different viewing angles $\theta=\pi/4$, $\pi/3$, $\pi/2$ and $3\pi/4$, respectively. For $\theta=\pi/3$, $\pi/2$ and $3\pi/4$, the peaks of these curves occur at $\epsilon_0 \approx 1$ MeV, and it seems that both the  high and the low energy range have smaller polarizations for $\theta> \pi/2$. This could be understood by the integration of $F_0$, since $F_0$ tends to have a relatively smaller value when $\epsilon_0$ is around $1$ MeV. \citet{Fan:2009} used an alternative method to study the similar configuration, and got the maximum polarization, which roughly agrees with that of  \citet{McNamara:2009} and our results. We conclude that the isotropic electrons reduce the initial polarization at large viewing angles. At small viewing angles, large polarization is mainly retained. The isotropic electrons can also induce the polarization for unpolarized initial beams, and the maximum degree of polarization is less than $15\%$.

Both \citet{McNamara:2009} and \citet{Krawczynski:2012} used the Monte-Carlo simulation to study the polarization in the case that unpolarized photons scattered by the electron beam in the jet. However, there is a discrepancy between them. In Section 3.4 of Krawcaynski (2012), it was shown that the polarization degree of inverse Compton emission of unpolarized radiation fields vanishes. However, the Lorentz factor of electrons was taken to be $10,000$ in the simulation of Krawcaynski. In the work of \citet{McNamara:2009}, the range of the Lorentz factor of electrons was taken from $1$ to $10,000$.
Based on our work, when a photon beam collides with an electron beam, there should be net polarization at certain viewing angles for the initially unpolarized photon beam. We showed that the contribution of electrons with Lorentz factor larger than $10$ can be ignored. We also showed that maximal $15\%$ polarization can be achieved for the Lorentz factor of electrons in the range $1 \sim 10$. Therefore, the discrepancy of the results of \citet{McNamara:2009} and \citet{Krawczynski:2012} can be explained by the different energy ranges of electrons.

\subsection{Thermal electrons}

\begin{figure}
\centering
  \plotone{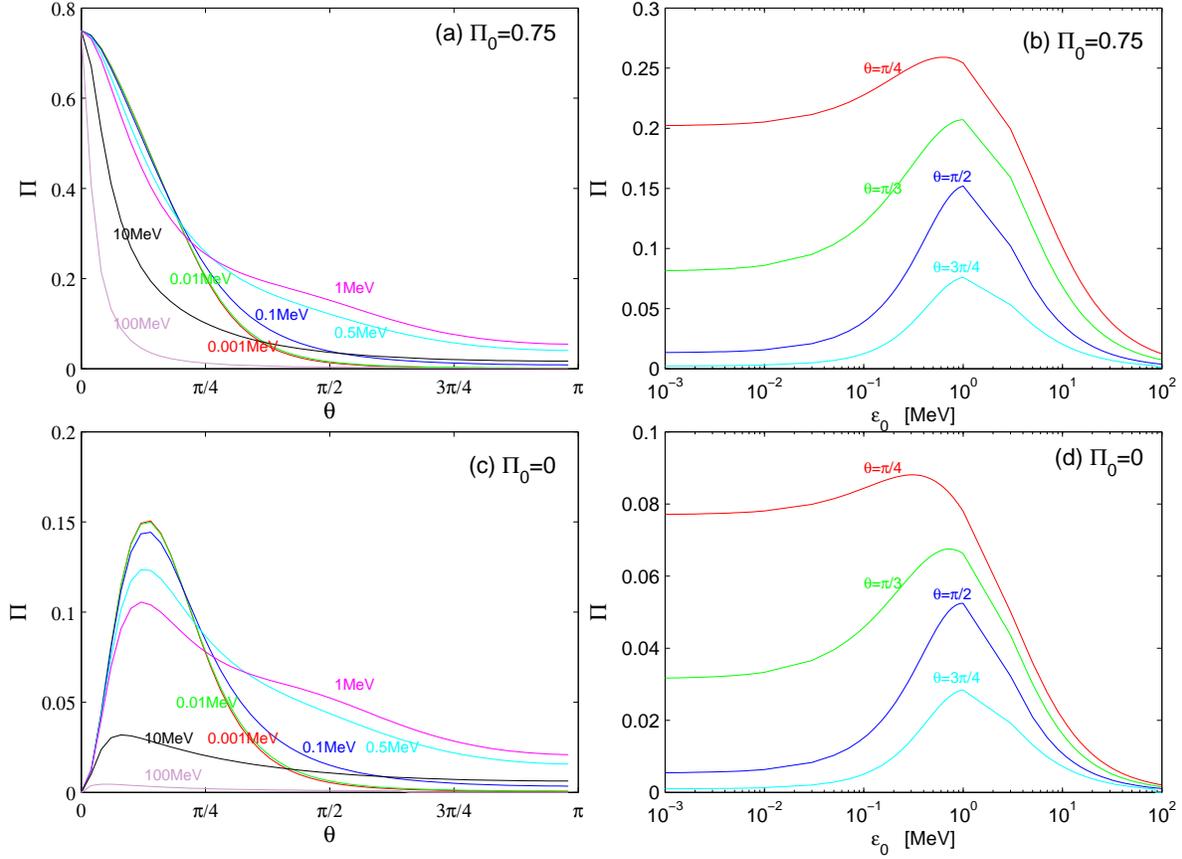}
  \caption{Polarization as a function of  $\epsilon_0$ and  $\theta$ for the thermal electrons. The thermal temperature is chosen to be $kT=200$ keV, where $k$ is the Boltzman constant. The top two panels describe the incident beam is polarized with  $\Pi_0=0.75$. The bottom two panels show the initially unpolarized case. Other settings are the same with that in Fig. \ref{fig:pipower}. } \label{fig:pithermal}
\end{figure}

Photons scattered by thermal electrons are also common in astrophysical processes. The thermal spectrum of electrons can be expressed as
\beq \label{eq:the}
{\cal N}(E) \propto \frac{E^2}{e^{E/kT}-1},
\eeq
where  $k=8.6 \times 10^{-5}$ eV/K is the Boltzman constant, $T$ is the temperature of electrons in unit of $K$ (Kelvin). $E$ is the kinetic energy of
 electrons, i.e., $E=(\gamma-1)m_e c^2=E_0-m_e c^2$. As a typical example, we choose $kT=200$ keV.
 In Fig. \ref{fig:pithermal}, the polarization for thermal electrons is plotted. The curves in Fig. \ref{fig:pithermal} are similar to that of Fig. \ref{fig:pipower}. The right two panels show that the polarization increases with energy up to around $1$ MeV, and then decreases. The maximal $\Pi$ is about one time smaller than that in the power law case. For instance, the polarization is about $40\%$ at $\theta \approx \pi/4$ and $\epsilon_0\approx 1$ MeV in the right top panel of Fig. \ref{fig:pipower}, while the polarization is about $25\%$ in the right top panel of Fig. \ref{fig:pithermal}. This indicates that  thermal electrons reduce the polarization of incident photons much significantly than electrons of power law distribution do. For the initially unpolarized case, the peak of the polarization curve appears around  $\pi/8$ (see the left bottom panel) for soft $\gamma$-rays. The right bottom panel shows that the final polarization caused by  thermal electrons is less than $8\%$, and this is difficult to be observed.

\section{Discussion and conclusions }\label{sec:DC}

Our formulae for the polarization of scattered photons are based on the quantum electrodynamics, and  cover both the Thomson limit and the KN limit. We mainly investigated the polarization in three scenarios, i.e., electrons at rest, isotropic electrons with a power law spectrum and thermal electrons.
Electrons at rest, namely the ``cold" electrons, can induce a completely linear polarization at the viewing angle $\theta \approx \pi/2$. Given an initially polarized beam, isotropic electrons will reduce the  polarization  significantly at large viewing angles for any incident energies, and the thermal electrons reduce the polarization more serious than the power law electrons. For the initially unpolarized beam, soft $\gamma$-rays can lead to $15\%$ polarization at viewing angles around $\pi/4$. For isotropic electrons,  one remarkable feature is that the polarization as a function of $\epsilon_0$ always peaks roughly at $1$ MeV, this is valid for both the thermal and power law cases. However, $\epsilon_0$ is not an observable variable. It is more convenient to plot the polarization as a function of $\epsilon_1$, which is difficult to realize in the setup for isotropic electrons. The scattering between a 1 MeV photon and an electron is always in the KN regime, since $\gamma_e \epsilon_0 > m_e c^2$. Roughly speaking, the scattered photon obtains the energy comparable to that of the relativistic electron. For the range $2 \lesssim \gamma  \lesssim 10$, the observed photon energy is in the band $1$ MeV $  \lesssim  \epsilon_1 \lesssim  3$ MeV, which is an expected range to observe this phenomenon.

The incident radiation are assumed to be monochromatic, but the outgoing radiation will have a spectral distribution. Suppose the electron distribution is a power law,  the total spectrum of ICS can be summarized as a broken power law \citep{Blumenthal:1970}
\begin{eqnarray} \label{eq:spectrum}
\frac{d N_{\rm tot}}{d t d \epsilon_1} \propto  \begin{cases}
  \epsilon_1^{-(p+1)/2}~~ &   {\rm Thomson \,\,\,limit}, \\
\epsilon_1^{-(p+1)} ~~&  {\rm KN \,\,\,limit}.
  \end{cases}
\end{eqnarray}
If the incident photons have a spectrum, one can add an extra factor in Equation (\ref{eq:spectrum}). Evidently, the final spectrum of ICS is independent of the spectrum of initial photons, and determined only by the spectrum of incoming electrons. The spectrum is steepened in the KN regime, the smoothly jointed spectrum here is of Band type.   Our analysis indicates that  a moderate linear polarization cannot exclude  ICS as an emission mechanism in GRBs. In the simulation, we take the energy range of electron to be $0.5$ MeV $< E_0 < 5$ MeV. Higher energy electrons only slightly affect the polarization. In the jet frame, the shocked electrons have a power law spectrum  with the minimal Lorentz factor $\gamma_{\rm min} \approx 610 \varepsilon_e \Gamma$, where $\epsilon_e$ is the fraction of the shock energy transferred  to electrons and $\Gamma$ is the Lorentz factor of the shocked fluid \citep{Sari:1998}.  Our calculation indicates that the integral of the range $\gamma>10$ contributes almost nothing to the polarization. Therefore, if the incident photons are scattered by shocked electrons, the initial polarization will be erased. In some GRBs, GeV photons were observed \citep{Abdo:2009,Ackermann2010}. If these GeV photons are of the SR origin, large linear polarization will be observed. If these photons are due to ICS, no significant polarization will be observed.

Our analysis is in the jet frame. One needs to consider the Lorentz transformation when the jet moves relatively towards the observer. For instance, the bulk Lorentz factor of the long GRB is about $250$ in the magnetic jet model \citep{Chang:2012}. In the jet frame, the polar angle of the outgoing direction can be as large as $180^{\circ}$. When the jet moves relativistically, the outgoing photons are in a cone with polar angle $\theta_{\rm obs}\lesssim  1/\Gamma$ in the observer frame. The polar angle $\theta$ in the jet frame is changed to $\theta_{\rm obs}$ in the observer frame via the relation
\beq
\cos \theta= \frac{\cos \theta_{\rm obs}- \bar{\beta} }{ 1- \bar{\beta} \cos \theta_{\rm obs}},
\eeq
where $\bar{\beta}$ corresponds to the bulk Lorentz factor $\Gamma$ of the jet. The degree of the polarization is  Lorentz invariant, i.e., $\Pi(\theta)= \Pi_{\rm obs}(\theta_{\rm obs})$. If $\beta$ is a constant, the observed $\Pi_{\rm obs}$ as a function of $\theta_{\rm obs}$ can be deformed easily from the left  panels in Fig. \ref{fig:pipower}. The net polarization should consider the radiation coming from different viewing angles, one needs to integrating over $\theta_{\rm obs}$ to get an averaged value \citep{Chang:2013}. As a result, the bulk motion of the jet strongly affects the observed polarization.

The observed polarization depends on the last scattering process in the optically thin region. In our setup, we assume that  photons are scattered only once by electrons.  However, interactions become complicate when the plasma is in the optically thick region. For instance, if the incident photon has energy above $1$ MeV, it could collide with the scattered photons to produce the electron-positron pair. If the optically thick condition always holds, the photon energy in the jet frame finally has an upper limit 1 MeV \citep{Beloborodov:2009be}. In many astrophysical conditions, the multiple scattering of photons may occur. The initial photon changes its polarization and momentum after one scattering process. The final  photons before the last scattering may form an axial symmetric distribution relative to the jet moving direction, or an isotropic distribution for that the optical depth is high enough. So, we should ask what is the polarization for photons with a given distribution scattered by isotropic electrons.
We find it difficult to give analytical formulae for this case. In our work, we take use of the coincidence that the kinematic equation and Stokes parameters are defined in the same frame of $xyz$ axes. However, this condition does not hold for the non-beamed incident photons. Because there is no one uniform coordinate system in which Stokes parameters of all incoming photons can be described. \citet{Nagirner:1993} used the $4-$vector to describe the polarization basis of the photon to overcome this difficulty.  Formulae of Compton scattering matrix for the scattering of radiation with given energy  by isotropic electrons were presented, and the relativistic kinematic equation which describes the Compton scattering process was formulated in the linear approximation \citep{Nagirner:1993,Nagirner:1994}. The interested reader are referred to their work.

We have considered the isotropic electrons. It is not difficult to generalize our analysis to certain conditions where the system has other geometries. For instance, one can consider the pencil beams of electrons, where electrons are injected along a small cone along the jet axis, by adding a factor $\exp\{- \alpha \theta_2^2\}$ to the integral ($\alpha$ is a constant). This will recover the results obtained by \citet{Begelman:1987}.
Although there is difficulty to obtain analytical formulae for the incident photons with a given distribution, it is not difficult to apply our result to the Monte Carlo simulation to obtain the numerical results. \citet{Krawczynski:2012} used the Monte Carlo simulation to study the polarization in the KN regime. The polarization of one ICS event was considered in such a way that  the Stokes vector was transformed from the plasma frame to the electron  rest frame, calculated by using the Fano matrix, and  transformed back to the plasma frame to obtain the results. Then, Monte Carlo simulations were used to perform the numerical calculations for isotropic electrons. In our analysis, one should build a frame for each incident photon, in which the $z'$ axis is set to be the incoming direction of the photon. This frame is only a spatial rotation of the jet frame. Then, the relation of Stokes parameters in these two frames can be found. In this way,  steps to do Lorentz transformations of all the parameters can be avoided.  So, our work may help to reduce the calculation in the Monte-Carlo simulation to study the case where photons are not beamed.

Compton scattering is an important radiation process for many astrophysical sources. Our study indicates that Compton scattering can induce the polarization of photons in a wide energy range, i.e, from radio to $\gamma$-ray. Thus, it enriches the construction of models with Compton scattering to explain the observed polarization. One of our results shows that moderate polarization can be induced when initially unpolarized photons pass through isotropic electrons. Another result indicates that the observed polarization depends on the initial photon energy with the fixed viewing angle. A further study is needed to investigate the relation between the degree of polarization and the observed photon energy.  However, the verification of this relation is an extremely challenging task with the current technology.

The widely accepted paradigm for AGN includes a massive black hole as the central engine, surrounded by an accretion disk and many fast-moving clouds near or far from the center \citep{Ghisellini:2012}. A dusty torus obscures the emission at the transverse direction of the jet. Some hot electrons may pervade the spatial region between clouds. The classification of AGN can be intrinsically due to different viewing angles. Blazars are a special class of AGN whose jets are oriented more or less towards the observer. Blazars can be classified into two categories: the BL Lac objects (BL Lacs) and the flat spectrum radio quasars (FSRQs), depending on the absence or presence of emission lines. A bright and powerful FSRQ, the source of 3C 454.3, has $\gamma$-ray emission detected for several years. The overall spectral energy distribution (SED) of this Blazar, if plotted in $\nu F_{\nu}$, shows a bimodal pattern \citep{Bonnoli:2011}. The first peak occurs between the radio and the optical, and can be interpreted as the synchrotron radiation directly due to its quick variability. The second high energy peak locates in the GeV band, the production of it can be due to the IC process of the seed synchrotron photons by the corresponding source electrons. According to this SED, one can estimate the polarization of the GeV photons by using our framework. Inversely, our work can be tested in the AGN phenomenon, although the experimental measurement of the polarization in such high energy is still in the far future.

In Seyfert galaxies with a bright core, there is an AGN. Seed photons from the accretion disk may collide with the hot electrons in the bulk region. The scattered photons can be observed at a certain viewing angle in the Type 1 Seyfert galaxy, or be obscured by the torus in the Type 2 Seyfert galaxy \citep{Ghisellini:2012}. The optical spectropolarimetry of the Seyfert 1 galaxy Fairall 51 indicated that the polarization is in the range from $5\%$ to $13\%$ \citep{Schmid:2001}. From Fig.3, the unpolarized photons, after colliding with isotropic electrons, can reach a maximal polarization degree $13\%$ at certain viewing angles. It is possible that such process may be one emission mechanism in the Type 1 Seyfert galaxy. Another feature of the emission is that the radiation has high and variable polarization in the radio and the optical range. An abrupt $90^\circ$ flip of the polarization orientation in the radio was observed after the gamma ray flare in November 2011 \citep{Orienti:2013}. In GRB 100826A, Yonetoku et al. also observed the $90^\circ$ rotation of the polarization orientation \citep{Yonetoku:2011}. The Compton scattering may play an important role in explaining such phenomenon. This is an interesting topic for the future investigation.

\begin{acknowledgments}
We are grateful to M. H. Li, X. Li and S. Wang for useful discussion. The work of Z. Chang and H. N. Lin has been funded by the National Natural
Science Fund of China (NSFC) under Grant No. 11075166 and No. 11375203. The work of Y.~G.~Jiang has been funded by NSFC under Grant No. 11203016 and No. 11143012.
\end{acknowledgments}


\begin{thebibliography}{}
\bibitem[Abdo et al., 2009]{Abdo:2009}Abdo, A. A., et al. 2009, Science, 323, 1688
\bibitem[Ackermann et al., 2010]{Ackermann2010}Ackermann, M., et al. 2010, ApJ, 716, 1178
\bibitem[Aharonian \& Atoyan, 1981]{Aharonian:1981}Aharonian, F. A., \& Atoyan, A. M. 1981, \apss, 79, 321
\bibitem[Befelman \& Sikora, 1987]{Begelman:1987}Begelman, M. C., \& Sikora, M. 1987, \apj, 322, 650
\bibitem[Beloborodov, 2010]{Beloborodov:2009be}Beloborodov, A.~M. 2010,  MNRAS, 407, 1033
\bibitem[Berestetskii et al., 1982]{Berest:1982}Berestetskii, V. B., Lifshitz, E. M., \& Pitaevskii, L. P. 1982, Quantum electrodynamics (Pergamon Press)
\bibitem[Blumenthal \& Gould, 1970]{Blumenthal:1970}Blumenthal, G. R., \& Gould, R. J. 1970, Rev. Mod. Phys, 42, 237
\bibitem[Bonnoli et al., 2011]{Bonnoli:2011}{Bonnoli, G., et al. 2010, \mnras, 410, 368}
\bibitem[Bonometto et al., 1970]{Bonometto:1970}Bonometto, S., Cazzola, P., \& Saggion, A. 1970 \aap, 7, 292
\bibitem[Bonometto \& Saggion, 1973]{Bonometto:1973}Bonometto, S., \& Saggion, A. 1973, \aap, 23, 9
\bibitem[Celotti \& Matt, 1994]{Celotti:1994}Celotti, A., \& Matt, G. 1994, \mnras, 268, 451
\bibitem[Chang et al., 2012]{Chang:2012}Chang, Z.,Lin H. N.,  \&  Jiang, Y. G. 2012, \apj,  759, 129
\bibitem[Chang et al., 2013]{Chang:2013}Chang, Z., Jiang, Y. G., \& Lin H. N. 2013, \apj, 769, 70
\bibitem[Fan, 2009]{Fan:2009}Fan, Y. Z. 2009, MNRAS, 397, 1539
\bibitem[Fano, 1957]{Fano:1957}Fano, U. 1957, Rev. Mod. Phys. 29, 47
\bibitem[Foley et al., 2011]{Foley:2011}Foley, S., et al. 2011, GCN Circular, 11771
\bibitem[Frail et al., 1998]{Frail:1998}Frail, D. A., Kulkarni, S. R., Bloom, J. S., \& Djorgovski, S. G. 1998, GCN Circ. 147
\bibitem[Ghisellini, 2012]{Ghisellini:2012}{Ghisellini, G., 2012, arXiv:1202.5949v1}
\bibitem[Ginzburg \& Syrovatskii, 1965]{Ginzburg:1965}Ginzburg, V. L.,  \& Syrovatskii, S. I. 1965, \araa, 3, 297
\bibitem[Ginzburg \& Syrovatskii, 1969]{Ginzburg:1969}Ginzburg, V. L.,  \& Syrovatskii, S. I. 1969, \araa, 7, 357
\bibitem[G\"{o}tz et al., 2009]{Gotz:2009}G\"{o}tz, D., et al. 2009, \apj, 695, L208
\bibitem[Kibble, 1960]{Kibble:1960}Kibble, T. W. B. 1960, Phys. Rev., 117, 1159
\bibitem[Krawczynski, 2012]{Krawczynski:2012}Krawczynski, H. 2012, \apj, 744, 30
\bibitem[Mandel \& Wolf, 1995]{Mandel:1995}Mandel, L., \& Wolf, E. 1995, Optical coherance and quantum optics, Cambridge
\bibitem[McNamara et al., 2009]{McNamara:2009}McNamara, A. L., Kuncic, Z., \& Wu, K. 2009, MNRAS, 395, 1507
\bibitem[Nagirner \& Poutanen, 1993]{Nagirner:1993}Nagirner, D. I., \& Poutanen, J., 1993, \aap, 275, 325
\bibitem[Nagirner \& Poutanen, 1994]{Nagirner:1994} Nagirner, D. I., \& Poutanen, J., 1994, \apspr, 9, 1
\bibitem[Orienti et al., 2013]{Orienti:2013}Orienti, M., et al. 2013, arXiv:1309.5286
\bibitem[Rybicki \& Lightman, 1979]{Rybicki:1979}Rybicki, G. B., \& Lightman, A. P. 1979,  Radiative Processes in Astrophysics (New York: Wiley)
\bibitem[Sari et al., 1998]{Sari:1998}Sari, R., Piran, T., \& Narayan, R. 1998, \apj, 497, L17
\bibitem[Schmid et al., 2001]{Schmid:2001}{Schmid , H. M., et al. 2001, \aap, 372, 59}
\bibitem[Westfold, 1957]{Westfold:1957}Westfold, K. C. 1957, \apj, 130, 241
\bibitem[Yonetoku et al., 2011]{Yonetoku:2011}{Yonetoku, D., et al. 2011, \apjl, 745, L30}
\end{thebibliography}
\end{document}